\documentclass[twocolumn]{aastex61}

\shorttitle{negative flare}
\shortauthors{Chelpanov et al.}
\begin{document}
\title{Oscillations accompanying a He\,\textsc{i} 10830\,\AA\ negative flare in a solar facula}
\correspondingauthor{A.A.~\surname{Chelpanov}}
\email{chelpanov@iszf.irk.ru}
\author{Andrei~\surname{Chelpanov}}
\affil{Institute of Solar-Terrestrial Physics
                     of Siberian Branch of Russian Academy of Sciences, Irkutsk, Russia}
\author{Nikolai~\surname{Kobanov}}
\affiliation{Institute of Solar-Terrestrial Physics
                     of Siberian Branch of Russian Academy of Sciences, Irkutsk, Russia}
\begin{abstract}

On September 21, 2012, we carried out spectral observations of a solar facula in the Si\,\textsc{i} 10827\,\AA, He\,\textsc{i} 10830\,\AA, and H$\alpha$ spectral lines. Later, in the process of analyzing the data, we found a small-scale flare in the middle of the time series. Due to an anomalous increase in the absorption of the He\,\textsc{i} 10830\,\AA\ line, we identified this flare as a negative flare.

The aim of this paper is to study the influence of the negative flare on the oscillation characteristics in the facular photosphere and chromosphere.

We measured line-of-sight (LOS) velocity and intensity of all the three lines as well as half-width of the chromospheric lines. We also used SDO/HMI magnetic field data. The flare caused modulation of all the studied parameters. In the location of the negative flare, the amplitude of the oscillations increased four times on average. In the adjacent magnetic field local maxima, the chromospheric LOS velocity oscillations appreciably decreased during the flare. The facula region oscillated as a whole with a 5-minute period before the flare, and this synchronicity was disrupted after the flare. The flare changed the spectral composition of the line-of-sight magnetic field oscillations, causing an increase in the low-frequency oscillation power.
\end{abstract}

\section{Introduction} \label{sec:intro}

Solar magnetic structures determine the development of most processes in the solar atmosphere. Of such structures, sunspots and faculae are the most ubiquitous. Faculae occupy notably greater areas than sunspots do, and they can be observed at the Sun even during low-activity periods. Due to their pervasiveness, faculae play an important role in the energy transport processes. One of the agent in the transport mechanisms is waves.

Studying oscillations and waves in faculae started in the 1960s \citep{Orrall65,Sheeley71,Deubner74} and continues intensely nowadays \citep{Khomenko08, Kobanov11SoPh}. Most of the works are based on intensity and line-of-sight velocity observations. The characteristics of oscillations are studied at different heights from the photosphere to the corona \citep{Kobanov2013SoPh284,KobanovChelpanov14}. \citet{Kostik13, Chelpanov15,Chelpanov16} studied the dependence of the oscillation periods on the magnetic field strength. Much rarer are the observations on periodic variations of the magnetic field strength in faculae \citep{Muglach95,KobanovPulyaev07,Strekalova16, Kolotkov17}. Usually, these studies are carried out for ``quiet periods of time" in the facula life cycle, i.e., in the absence of flare phenomena in them.

Numerous studies of oscillations related to strong solar flares in active regions with sunspots were carried out in other bands, e.g., optical, radio, EUV, or X-ray \citep{NakariakovMelnikov09, Mariska10, Nakariakov10, Kallunki12, Balona15, Kolotkov15}. Most flares in active regions occur close to the magnetic inversion line, i.e., outside sunspots' umbrae. The strength and configuration of the magnetic field in the vicinity of such areas are similar to those in the spotless facular regions, and one can assume that oscillation characteristics in such structures have much in common. Quasi-periodic pulsations (QPP) are observed during flares in a broad range of wavelengths and atmospheric heights: the H$\alpha$ line (e.g., \citet{Srivastava08, YangXiang}), EUV \citep{Su12}, XR \citep{Zimovets10,Chowdhury15, LiZhang}. One can consider three main scenarios concerning relations of flares with oscillaitons. The first: flares are the sources for oscillaitons \citep{Mariska06}. \citet{LiZhang} showed that the oscillation period increases with the increase in the X-ray photon energy. The second scenario: high-amplitude oscillations cause reconnection of the closely located magnetic filaments triggering a flare \citep{Inglis09}. \citet{NakariakovZimovets11} showed that slow magnetoacoustic waves can propagate along magnetic arcade axes and cause their periodic reconnection during a flare. \citet{Sych2009}, while studying AR 10756, found that magnetoacoustic 3-minute waves propagate by the magnetic channels from the umbra to the periphery of the AR, where they trigger flares. According to them, an increase in 3-minute oscillation amplitude in the umbral chromosphere may serve as a flare precursor. The third scenario: a flare occurs independently of the oscillations in the studied object only modulating some of the observed oscillation parameters. These scenarios can also combine with each other.

QPPs are commonly registered in the hard X-ray and microwave ranges during strong flares. Periods of 10 to 40 seconds are usually registered \citep{Inglis16}. Based on RATAN-600 microflare observations, periods of 1.4 and 0.7\,s were found in the radio emission intensity and polarization signals, respectively \citep{Nakariakov2018}. \citet{Simoes15} found that 28 out of 35 researched x-flares showed QPPs. Some authors relate changes in the period with the flare phases. \citet{Hayes16} showed that a 40-second period is typical of the impulse phase, while a period of 80 seconds is observed during the gradual phase. \citet{Dennis17} observed the period change from 25 to 100 seconds in the course of a flare development. \citet{Pugh17} found no dependence of the QPP period on such active region characteristics as bipole separation distance or average magnetic field strength. One rarely observes periods of 2--3 minutes in the soft and hard X-ray oscillations. Nevertheless, analysing an X-class flare, \citet{Li15} registered 4-minute period oscillations in a wide range of wavelengths from the X-ray and EUV to the microwave range. \citet{Milligan17} found that during flares, Ly{$\alpha$} (1216\,\AA) shows an increased 3-minute period power in the whole-disk radiation. A similar increase in the power was registered in the 1600 and 1700\,\AA\ channels. \citet{Zhang17} suggested that surge-like oscillations of a 6.5-minute period above light sunspot bridges may be caused by shock p-mode waves leaking from the photosphere. The review given above includes but a fraction of the massive volume of studies dedicated to QPPs. As \citet{McLaughlin18} underlined, waves in flares can be used to diagnose flare processes. We believe that to this end, one should study how flares influence the regime of the already existing oscillations, which are abundant in the solar atmosphere. This will help distinguish the oscillations generated by the flare from those that themselves might cause the magnetic reconnection.


In this paper, we study the characteristics of photospheric and chromospheric oscillations within an isolated facular region, where a small-scale flare occurred during the observations. Due to an anomalous (up to 25\%) absorption increase in the He\,\textsc{i} 10830\,\AA\ line, we identified this flare as a \textit{negative flare} \citep{Kobanov18}. Such solar events are rarely observed\citep{Henoux, Xu16}, and a study of oscillation processes related to this flare, which is the aim of the paper, is of great interest. To the best of our knowledge, for such events has been presented no observational data that includes oscillations of Doppler velocity, intensity, profile half-width, and LOS magnetic field strength.

\section{Instrument and methods}

The observations were carried out with the Horizontal Solar Telescope at Sayan Solar Observatory. The telescope is equipped to measure line-of-sight (LOS) velocity and magnetic fields \citep{Kobanov01, Kobanov13}. The telescope is located 2000\,m above sea level; it is elevated 6\,m above the ground. The telescope is equipped with flat 800-mm ceolostat mirrors and a spherical 900-mm main mirror, whose focal length is 20\,m; the mirrors are made of sitall glass, which provides resistance to temperature changes. A photoelectric guide compensates the object movements due to the rotation of the sun; it is also used for automated scanning of an observed object. The combination of these optical and mechanical equipment provides the precision of the object tracking and scanning of under 1$''$.

In the spectral observations, the field of view covers a 1.5$\times$60$''$ area. Camera's one pixel corresponds to 0.24$''$ distance on the solar disk and 5 to 16\,m\AA\ of the spectrum. The data processing was performed using IDL (Interactive Data Language).

In addition to our ground-based observations, we used data from the Helioseismic and Magnetic Imager (HMI, \citet{sdohmi}) onboard the Solar Dynamics Observatory (SDO).

We measured velocity signals using the lambdameter technique \citep{rayrole}: the displacements of the line profiles were determined based on equal-intensity positions of two segments of the line wings with a given fixed distance between them. The positive values of the LOS velocity correspond to the movements towards the observer. The spectral line profile half-width was determined as the width of the profile at the half intensity level between the line core and the adjacent continuum. The Fast Fourier Transform algorithm was used to analyse the the oscillation spectral composition. The frequency filtration of the signals was performed with the use of the Morlet wavelet of the sixth order.

\section{Observation data}

\begin{figure}
\centerline{
\includegraphics[width=5cm]{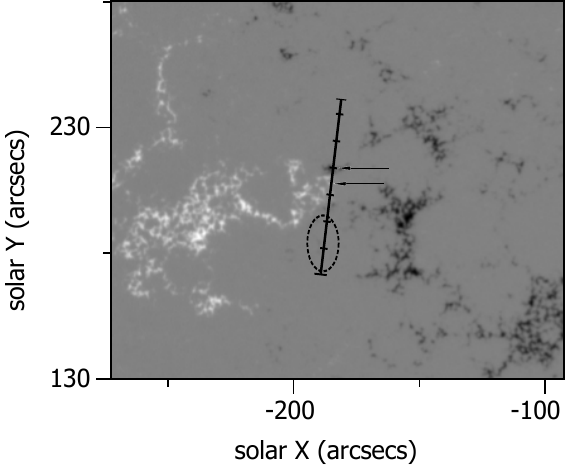}
}
\caption{The location of the spectrograph slit in the magnetic field map of the observed facula. The dashed oval shows the negative flare location. The two horizontal arrows show the locations of the local magnetic maxima that we used in the analysis.}
\label{fig:1}
\end{figure}

The facular region that we observed on September 21, 2012, 01:37 through 03:17\,UT (100 minutes), located close to the centre of the disk at N19E13. The flare was of a low power (under B2), and, to our knowledge, it is registered in no flare catalog. The start and end moments of the flare in the negative flare location were determined based on the H$\alpha$ light curve. The X-ray emission in the 6--12\,keV range, however, increased two minutes earlier. A more detailed description of the flare can be found in \citet{Kobanov18}. A B\,8.7-class flare occurred in this facula ten hours later. The magnetic field map of the region shows a bipolar structure; the slit was placed approximately at the polarity inversion line (Figure\,\ref{fig:1}). We carried out the observations in three spectral lines simultaneously: Si\,\textsc{i} 10827\,\AA, He\,\textsc{i} 10830\,\AA, and H$\alpha$. The cadence was 1.5\,s.

\section{Results and Discussion}

\begin{figure}
\centerline{
\includegraphics[width=8.5cm]{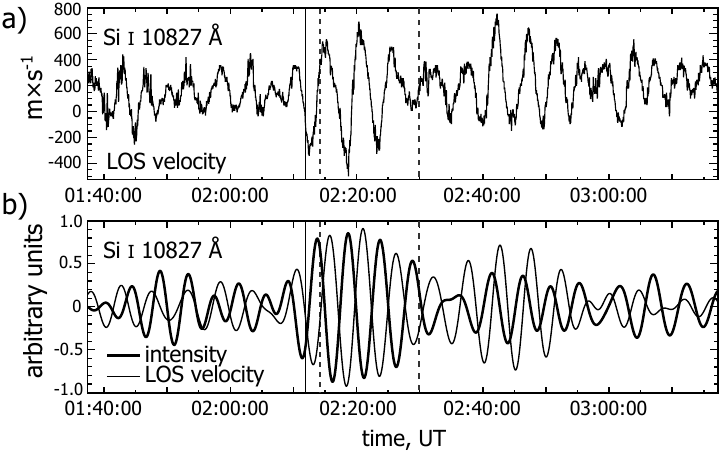}
}
\caption{Line-of-sight velocity and intensity signal of the Si\,\textsc{i} 10827\,\AA\ line in the negative flare location at the 10$''$ slit position. a) raw LOS velocity signal b) velocity and intensity Si\,\textsc{i} signals filtered in the 3.0--4.0\,mHz range. The \textit{y}-zero marks are set as the mean values of the series. Vertical dashed lines here and in the other figures show the start and end of the negative flare based on the H$\alpha$ light curve. The vertical solid line marks the start time based on the X-ray 6--12\,keV RHESSI channel.}
\label{fig:10-vel-int-SiHe}
\end{figure}

\begin{figure}
\centerline{
\includegraphics[width=8.5cm]{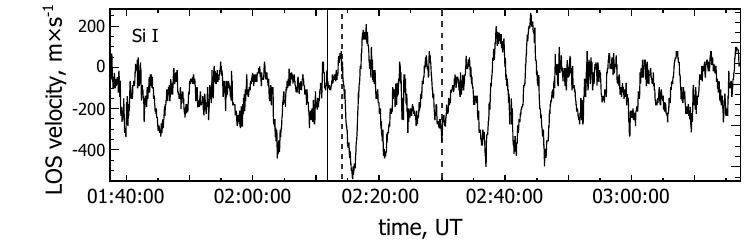}
}
\caption{Raw line-of-sight velocity signal of the Si\,\textsc{i} 10827\,\AA\ line in the location of the magnetic field maximum at the 34$''$ slit position. The vertical-axis zero velocity is set to the average value over the series.}
\label{fig:34-vel-Si}
\end{figure}

Spectral observations of photospheric oscillations were carried out in the Si\,\textsc{i} 10827\,\AA\ line. According to \citet{Bard08}, it forms at 540\,km, which corresponds to the temperature minimum. Of all the parameters observed in the negative flare location (10$''$ slit position), the LOS velocity signal shows the most distinct 5-minute oscillations typical of the photosphere (Figure\,\ref{fig:10-vel-int-SiHe}a). Two oscillation trains with an amplitude increased 2--2.5 times stand out in the non-filtered velocity signal: the first one coincides with the flare, while the second follows 15 minutes later. The phase lag between the intensity and LOS velocity signals was $\sim$180$^\circ$ during the flare, and $\sim$90$^\circ$ after the flare (Figure\,\ref{fig:10-vel-int-SiHe}b). Such phase relations indicates that during the flare, the oscillation regime in the photosphere did not correspond to standing acoustic waves. After the flare, a standing acoustic wave regime settled. Note that at the 34$''$ slit position in a local magnetic field maximum, a similar two-train sequence is seen (Figure\,\ref{fig:34-vel-Si}).

\begin{figure}
\centerline{
\includegraphics[width=9cm]{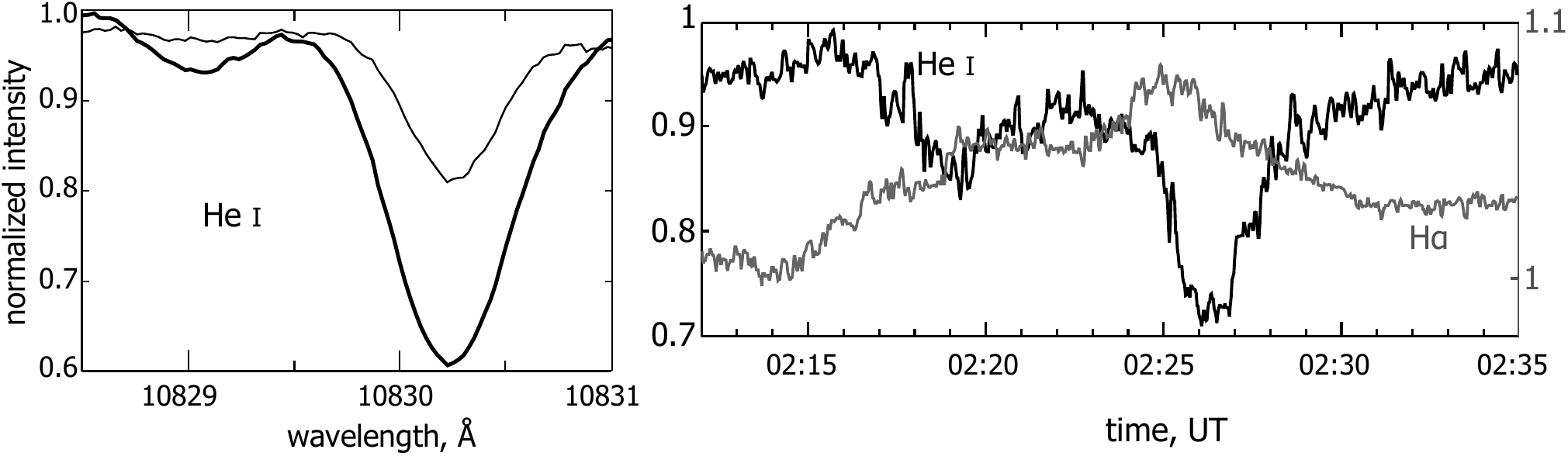}
}
\caption{\textit{Left:} the He\,\textsc{i} 10830\,\AA\ line profile during the negative flare (thick line) plotted against an undisturbed facular profile (thin); the dip at 10829\,\AA\ is the blue component of the He\,\textsc{i} triplet; \textit{right:} the light curves of the He\,\textsc{i} and H$\alpha$ lines during the flare. The plots are normalized to the pre-flare line core intensities.}
\label{fig:profile}
\end{figure}

We observed chromospheric oscillations in the He\,\textsc{i} 10830\,\AA, and H$\alpha$ 6563\,\AA\ lines. They form at a height of 1500--2000\,km \citep{Avrett94, Vernazza81, Leenaarts12}. At the spectrograph slit, the flare was localized between the 0$''$ and 20$''$ positions with its maximum at the 10$''$ position. In the He\,\textsc{i} line, the flare manifested itself as a darkening, which reached 25\% in the line core (Figure\,\ref{fig:profile}). At the same location and time, the H$\alpha$ core intensity increased 8\%.

\begin{figure}
\centerline{
\includegraphics[width=8.5cm]{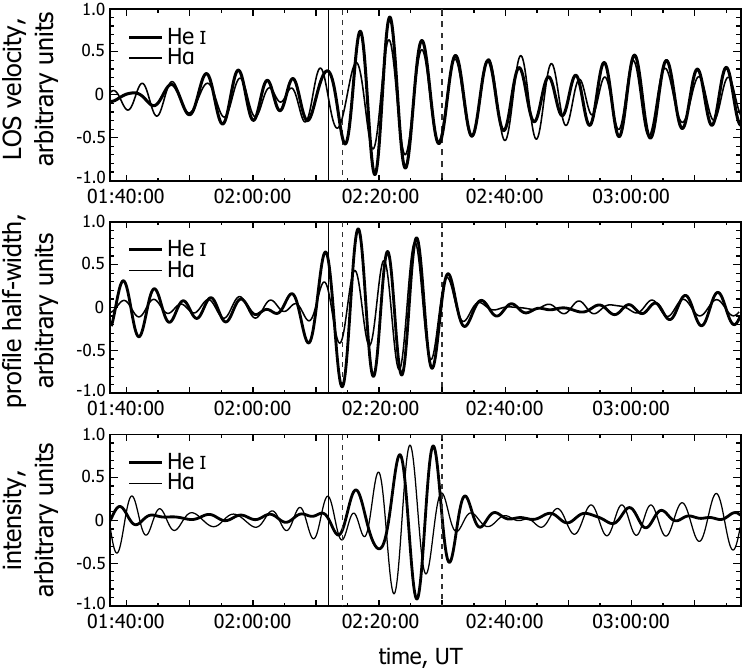}
}
\caption{Velocity, profile half-width, and intensity signals of the chromospheric lines in the flare location at the 10$''$ slit mark filtered in the 3.0--4.0\,mHz frequency range (5-minute oscillations).}
\label{fig:10-vel-hw-5min}
\end{figure}

\begin{figure}
\centerline{
\includegraphics[width=8.5cm]{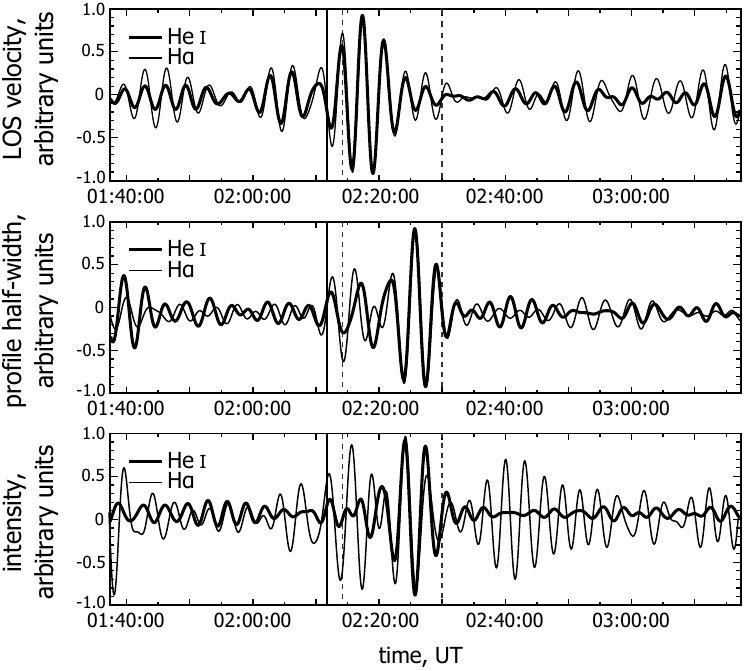}
}
\caption{Velocity, profile half-width, and intensity signals of the chromospheric lines in the flare location at the 10$''$ slit mark filtered in the 5.1--5.9\,mHz frequency range (3-minute oscillations).}
\label{fig:10-vel-hw-3min}
\end{figure}

Chromospheric oscillations in the negative flare location are represented by the LOS-velocity, profile half-width, and intensity of the H$\alpha$ and He\,\textsc{i} 10830\,\AA\ lines. To reveal the phase-time relationships between these signals, we filtered them in two frequency ranges: 3.0--4.0 and 4.5--5.5\,mHz. The resulting plots are presented in Figures\,\ref{fig:10-vel-hw-5min}\,and\,\ref{fig:10-vel-hw-3min}. One can see that during the flare (whose start and end times are marked with vertical lines) the oscillation amplitude increased in all the considered signals. Doppler velocity 5-minute oscillation amplitude of the the H$\alpha$ and He\,\textsc{i} 10830\,\AA\ lines increased three-fold. The half-width oscillations in this range increased 4--6 times.

The amplitude of the Doppler velocity 3-minute oscillations in these lines increased 4 times, and that of the half-width oscillation increased 4--5 times. Maximum of the half-width oscillations is about 8 minutes later after the Doppler velocity oscillations (Figure\,\ref{fig:10-vel-hw-3min}). One may suggest that this great increase in the oscillation amplitudes in the negative flare location in caused by coronal QPPs in the X-ray emission.


\begin{figure}
\centerline{
\includegraphics[width=8cm]{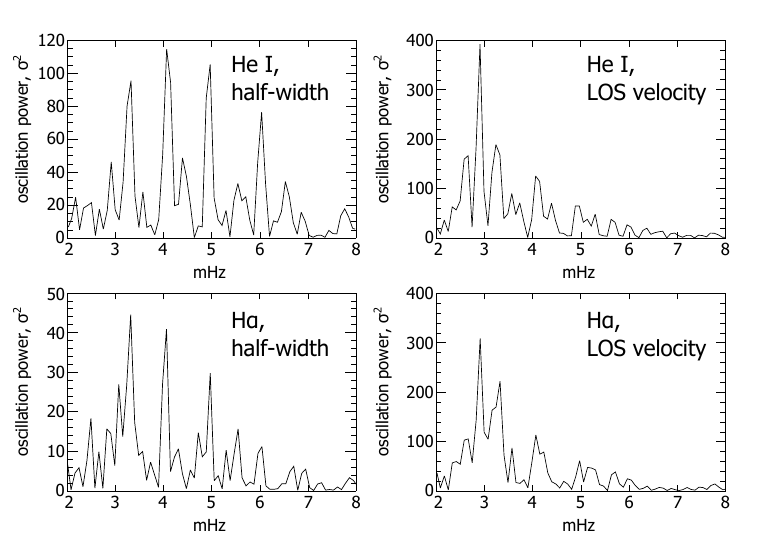}
}
\caption{Chromospheric line profile half-width and LOS velocity oscillation power spectra in the location of a facular local magnetic field maximum at the 34$''$ slit point.}
\label{fig:33spectra}
\end{figure}

\begin{figure}
\centerline{
\includegraphics[width=8.5cm]{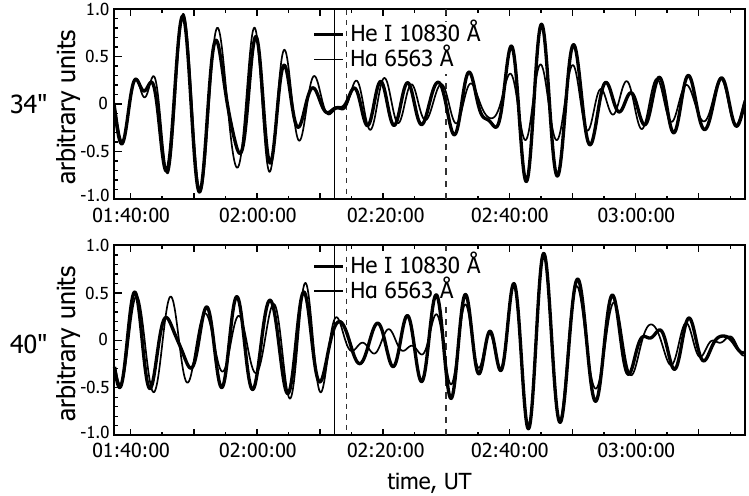}
}
\caption{Line-of-sight velocity signals of the chromospheric lines in local magnetic field maxima at the 34$''$ and 40$''$ spectrograph slit positions filtered in the 3.0--4.0\,mHz range.}
\label{fig:33-vel-HeHa}
\end{figure}

We also studied LOS velocity and profile half-width oscillations of the He\,\textsc{i}  and H$\alpha$ lines in one of the local magnetic field maximum (position 34$''$). LOS velocity oscillation power spectra in Figure\,\ref{fig:33spectra} show that 5-minute oscillations prevail, while profile half-width oscillation spectra show a more complex multiple-peak structure. To analyse the behaviour of the 5-minute oscillations, we filtered the signals in the 3.5$\pm$0.5\,mHz range. Like the photospheric oscillations (Figure\,\ref{fig:10-vel-int-SiHe}), the chromospheric signals demonstrate two oscillation trains. LOS velocity oscillations in the He\,\textsc{i} and H$\alpha$ lines show a decrease in the amplitude during the flare (Figure\,\ref{fig:33-vel-HeHa}). The time interval between the trains is 30 minute, and the first chromospheric oscillation train is 15 minutes ahead of its photospheric analogue. Figures\,\ref{fig:10-vel-hw-5min}, \ref{fig:10-vel-hw-3min}, and\,\ref{fig:33-vel-HeHa} show a clear synchronicity between the  He\,\textsc{i} and H$\alpha$ LOS velocity signals. In our opinion, this indicates that the vertical plasma oscillations are virtually simultaneous in the observed height range.

\begin{figure}
\centerline{
\includegraphics[width=8.5cm]{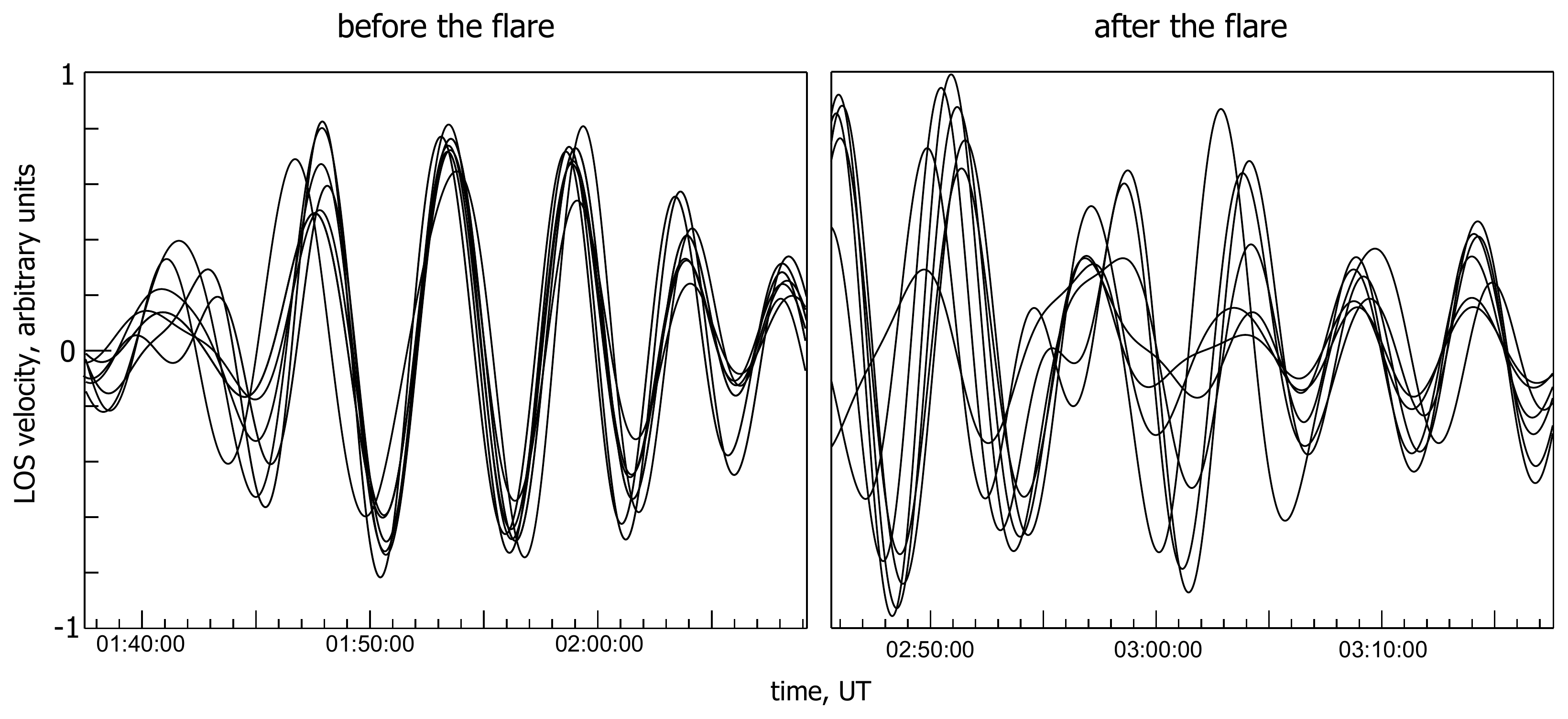}
}
\caption{Filtered 5-minute LOS velocity oscillations of the He\,\textsc{i} line in several locations along the slit before and after the flare.}
\label{fig:sync}
\end{figure}

A part of the facula adjacent to the negative flare location with widened chromospheric line profiles occupied the 25$''$--50$''$ interval at the spectrograph slit. The local maxima of the magnetic field situated in the same region. We divided this distance by seven segments, then filtered in the 3.5$\pm$0.5\,mHz range the He\,\textsc{i} LOS velocity signal in each segment. The left panel in Figure\,\ref{fig:sync} shows these signals in the pre-flare interval, and the right panel shows them after the flare. The signals in the left panel look synchronized as opposed to those in the right panel. This indicates that before the flare, the facula oscillated as a whole, and after the flare this synchronicity was disrupted. This may have resulted from the geometrical deformation of the facula coronal loop system caused by the flare. Such deformation may lead to changes in the oscillation spectral composition and amplitude-phase relations between different modes \citep{NakariakovPuls2018}.

\begin{figure}
\centerline{
\includegraphics[width=5cm]{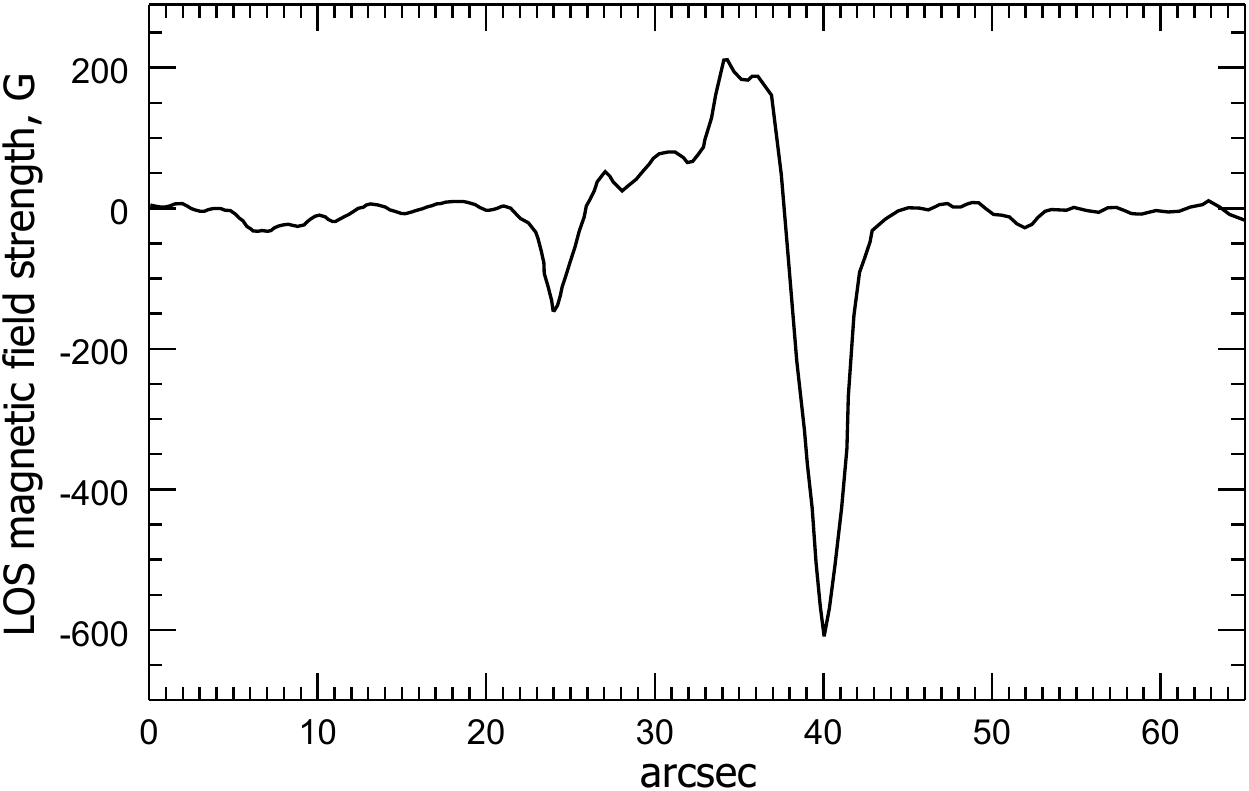}
}
\caption{Distribution of the magnetic field line-of-sight strength along the spectrograph slit before the flare at 02:10.}
\label{fig:mf-slit}
\end{figure}

\begin{figure}
\centerline{
\includegraphics[width=7.5cm]{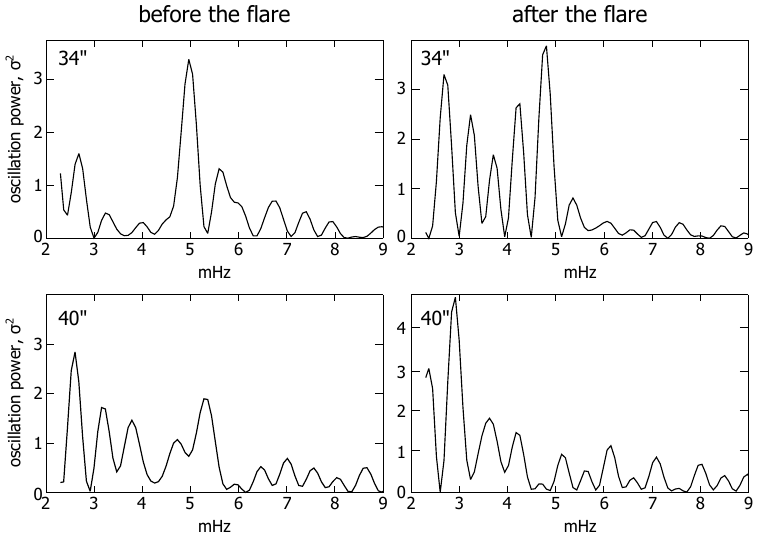}
}
\caption{Magnetic field oscillation spectra in the magnetic field local maxima (34$''$ and 40$''$ slit positions) before and after the flare.}
\label{fig:mf_spectra}
\end{figure}

Based on the HMI 6173\,\AA\ line data, we analysed the magnetic field line-of-sight oscillations in the magnetic strength maxima locations: 34$''$ and 40$''$ slit positions. Figure\,\ref{fig:mf-slit} shows the LOS magnetic field strength along the spectrograph slit. The local maxima at the 34$''$ and 40$''$ positions are of different magnetic polarity. The spectra in Figure\,\ref{fig:mf_spectra} show that the oscillation power of the LOS magnetic field increased after the flare in both locations in the 2.5--5.0\,mHz range. Note that in the 34$''$ position, the 5\,mHz peak dominates, and in the 40$''$ position the 3\,mHz peak dominates.

\citet{Emslie82} suggested that a part of the energy from the flare primary reconnection site in the corona may be transported to the lower atmosphere by Alfv{\'e}n waves, thus providing heating in the temperature minimum region. \citet{Fletcher08} developed this scenario showing that Alfv{\'e}n waves may result in the electron acceleration that causes quasi-periodic pulsations. They point out that, particularly, a torsional component is generated as a result of a reconnection in the twisted magnetic field.

Line profile half-width oscillations  can give an indirect observational evidence for torsional Alfv{\'e}n waves in the chromosphere \citep{Jess09,Kobanov17}. The increase in the half-width amplitudes may result from Alfv{\'e}n waves. In our observations, a sharp 5-fold increase in the half-width oscillation amplitude occurred during the flare, which may serve as an argument for the Alfv{\'e}n wave presence resulting from the flare.

In this context, three-minute oscillations are especially interesting, because they are not typical of faculae, as oppose to five-minute oscillations, which are normally observed in faculae \citep{Kobanov17}. In our data, 3-minute oscillations show high amplitudes only during the flare, and they cease right after it.

On the other hand, the intensity signals show clear 3-minute oscillations, whose power repeats that of the half-width oscillations (Figures\,\ref{fig:10-vel-hw-3min}), though torsional Alfv{\'e}n waves, being non-compressional, do not produce intensity oscillations. This suggests that the half-width oscillations that we observe are not the torsional Alfv{\'e}n waves.

Another possibility to generate 3-minute oscillations in the chromosphere is related to impulse movements in a gravitationally stratified environment \citep{ChaeGoode15,Kwak16}. Earlier, \citet{Botha11} emphasized and important role of the chromosphere resonator in the three-minute oscillation formation. As \citet{Kobanov18} noted, a series of impulse movements (up to 10--12\,km\,s$^{-1}$) followed the studied flare. In the 02:15--02:30 interval, the 3-minute oscillation period slightly increases over time, which agrees with \citet{Kwak16}. We cannot explain the time lag between the amplitude increases in the Doppler velocity and half-width signals.

Concluding, one can say that even a small flare (weaker than B2 in this case) is able to produce substantial perturbations in the oscillation regime of the lower solar atmosphere down to the temperature minimum level.

It is difficult to estimate whether the revealed features of the oscillation regime are characteristic of exclusively negative flares or common for all small flares. The studies on small flare-related oscillations in the lower atmosphere are scarce. For example, \citet{Leiko15} studied a microflare in the H$\alpha$ line and observed 3- and 5-minute oscillations throughout the whole 21-minute time series. They noted that 12 minute before the flare, the oscillation amplitude briefly increased four times.

Note that in the stellar flare studies, dips in light curves (which may be classified as negative flare) are observed more often than at the Sun \citep{Grinin1976, Rodono1979, Ventura1995, Leitzinger14}. This, and the fact that the spectral lines in which we observed are widely used in stellar observations make us hope that these results will turn out to be useful for the stellar astrophysics as well.

\section{Conclusion}

The studied facular flare whose feature was an anomalous (up to 25\%) increase in the He\,\textsc{i} 10830\,\AA\ line absorption caused a well-pronounced modulation of the main parameters oscillations in the lower layers of the solar atmosphere down to the photosphere. This influence of the flare considerably differs in various elements of the facula.

In the flare location, the oscillation amplitudes increased both in the upper photosphere (the Si\,\textsc{i} 10827\,\AA\ line) and in the chromosphere (the He\,\textsc{i} 10830\,\AA\ and H$\alpha$ lines). The phase lag between the intensity and LOS velocity photospheric signals was $\sim$180$^\circ$ during the flare, and $\sim$90$^\circ$ after the flare. Such phase relations indicates that during the flare, the oscillation regime in the photosphere did not correspond to standing acoustic waves. After the flare, a standing acoustic wave regime settled.

In the chromosphere in the location of the negative flare, the oscillation amplitude increased during the flare in both spectral lines in all the observed parameters: LOS velocity (increased 3--3.5 times), intensity (4--5 times), and profile half-widths (4--6 times). In the 3-minute range, the maximum of the profile half-width variations of the chromospheric lines lags $\sim$8 minutes behind the Doppler velocity signal.

In the local magnetic field maximum, the 5-minute oscillations prevail in the chromospheric velocity signals. They show two oscillation trains 30 minutes apart. The photospheric oscillation train follows the chromospheric one 15 minutes after (see Figures\,\ref{fig:34-vel-Si},\,\ref{fig:33-vel-HeHa}). Note that during the flare, 5-minute chromospheric oscillations abruptly decrease in amplitude.

In the 5-minute range, the He\,\textsc{i} line LOS velocity signals along the facula show simultaneous variations before the flare: the facula oscillates as a whole in this time interval. After the flare, the velocity oscillations are out of phase. This may have resulted from the geometrical deformation of the facula coronal loop system caused by the flare. Such deformation may lead to changes in the oscillation spectral composition and amplitude-phase relations between different modes.

Immediately after the flare, the oscillation power of the LOS magnetic field strength in two local magnetic field maxima increased in the 2.5--5.0\,mHz range. The magnetic field of the positive magnetic polarity showed a 5\,mHz dominant peak, while a 3\,mHz peak dominated in the negative polarity maximum.

We believe that the revealed high sensitivity of the oscillation characteristics in the lower solar atmosphere to small flares extends the possibilities of the wave diagnostics.

\acknowledgments

The research was partially supported by the Projects No.\,16.3.2 and 16.3.3 of ISTP SB RAS. Spectral  data on the chromosphere were recorded at the Angara Multiaccess Center facilities at ISTP SB RAS. We acknowledge the NASA/SDO science team for providing the magnetic field data. We are grateful to the anonymous referee for the useful suggestions.

\end{document}